\documentclass[12pt,a4,fleqn]{article}

\usepackage{epsfig}
\usepackage{latexsym}
\usepackage{amsmath}
\usepackage{amsfonts}
\usepackage{fancyhdr}
\def \beq {\begin{equation}}
\def \eeq {\end{equation}}
\def \bea {\begin{eqnarray}}
\def \eea {\end{eqnarray}}
\def \beqas {\begin{eqnarray*}}
\def \eeqas {\end{eqnarray*}}
\def \bse {\begin{subequations}}
\def \ese {\end{subequations}}

\newcommand{\bc}{\begin{center}}
\newcommand{\ec}{\end{center}}
\newcommand{\n}{\noindent}
\newcommand{\be}{\begin{equation}}
\newcommand{\ee}{\end{equation}}
\newcommand{\bear}{\begin{eqnarray}}
\newcommand{\eear}{\end{eqnarray}}

\begin{document}
\thispagestyle{fancy}\markright{\it {\footnotesize Mathematical and Computational Applications\\
\copyright Association for Scientific Research}}

\bc{\bf A CLASS OF CHARGED RELATIVISTIC SPHERES}\ec

\bc{ K.Komathiraj$^{1,2}$ and S.D.Maharaj$^{1}$\\
$^1$Astrophysics and Cosmology Research Unit, School of Mathematical
Sciences, University of KwaZulu-Natal, Private Bag X54001,  Durban
4000, South Africa.\\
$^2$Permanent address: Department of Mathematical Sciences, South Eastern University,
Sammanthurai, Sri Lanka.\\
maharaj@ukzn.ac.za, komathiraj@seu.ac.lk}\ec

\noindent
{\bf Abstract}- We find a new class of exact solutions to the
Einstein-Maxwell equations which can be used to model the interior
of charged relativistic objects. These solutions can be written in
terms of special functions in general; for particular parameter
values it is possible to find solutions in terms of elementary
functions. Our results contain models found previously for
uncharged neutron stars and charged isotropic spheres. 

\noindent
\bc{\bf 1. INTRODUCTION}\ec

The Einstein-Maxwell system of field equations are applicable in
modelling relativistic astrophysical systems. We need to generate
exact solutions to these field equations to model the interior of a
charged relativistic star that should be matched to the
Reissner-Nordstrom exterior spacetime at the boundary. A general
treatment of nonstatic spherically symmetric solutions with
vanishing shear was performed by Wafo Soh and Mahomed \cite{1} using
symmetry methods. The matching of nonstatic charged perfect spheres
to the Reissner-Nordstrom exterior was considered by Mahomed
\textit{et al}. \cite{2} who showed that the Bianchi identities
restrict the number of solutions. Particular models generated can be
used to model the interior of neutron stars as demonstrated by
Tikekar \cite{Tik}, Maharaj and Leach \cite{MaLe} and Komathiraj and
Maharaj \cite{KoMa}. Charged spheroidal stars have been widely
studied by Sharma \textit{et al}. \cite{ShMuMa} and Gupta and Kumar
\cite{GuSh}. There exist comprehensive studies of cold compact
objects by Sharma \textit{et al}. \cite{ShKaMu}, analysis of strange
matter and binary pulsar by Sharma and Mukherjee \cite{ShMu} and
quark-diquark mixtures in equilibrium by Sharma and Mukherjee
\cite{ShaMu}, in the presence of the electromagnetic field. Charged
relativistic matter is important in the modelling of core-envelope
stellar systems as demonstrated by Thomas \textit{et al}.
\cite{ThRa}, Tikekar and Thomas \cite{ThTi} and Paul and Tikekar
\cite{PaTi}. The recent treatment of Thirukkanesh and Maharaj
\cite{ThMa} deals with charged anisotropic matter with a barotropic
equation of state which is consistent with dark energy stars and
charged quark matter distributions.


The exact solution of Tikekar \cite{Tik} is spheroidal in that the
geometry of the spacelike hypersurfaces generated by $t$=constant
are that of a 3-spheroid. This condition of a spheroid helps to
mathematically interpret the solution since it provides a
transparent geometrical interpretation. On physical grounds we
find that this solution can be applied to model superdense stars
with densities of the order $10^{14}$  g cm$^{3}.$ The physical
features of the Tikekar model are therefore consistent with
observation, and consequently it attracts the attention of several
researches as a realistic description of the stellar interior of
dense objects. This solution was extended by Komathiraj and
Maharaj \cite{KoMa} to include the electromagnetic field, with
desirable physical features.  In this paper we show that a wider
class of solutions to the Einstein-Maxwell system is possible by
adapting the form of the gravitational potentials. Our intention
is to obtain simple forms for the solutions that are physically
reasonable and may be used to model a charged relativistic sphere.


\noindent
\bc{\bf 2. SPHERICALLY SYMMETRIC SPACETIME}\ec

The metric of  static spherically symmetric spacetimes in
curvature coordinates can be written as
 \be \label{eq:1}
ds^{2}=-e^{2\nu(r)}dt^{2}+e^{2\lambda(r)}dr^{2}+r^{2}(d\theta^{2}+\sin^{2}\theta
d\phi^{2})
\ee
where $\nu(r)$ and $\lambda(r)$ are two arbitrary
functions. For charged perfect fluids the Einstein-Maxwell system
of field equations is given by
 \bse\label{eq:2}
 \bea\label{eq:2a}
\frac{1}{r^{2}}(1-e^{-2\lambda})+\frac{2\lambda^\prime}{r}e^{-2\lambda}&=&\rho+\frac{1}{2}E^{2}\\
\label{eq:2b}\frac{-1}{r^{2}}(1-e^{-2\lambda})+\frac{2\nu^\prime}{r}e^{-2\lambda}&=&p-\frac{1}{2}E^{2}\\
\label{eq:2c}e^{-2\lambda}\left(\nu^{\prime\prime}+{\nu^\prime}^2+\frac{\nu^\prime}{r}-\nu^\prime\lambda^\prime-\frac{\lambda^\prime}{r}\right)&=&p+\frac{1}{2}E^{2}\\
\label{eq:2d}\sigma&=&\frac{1}{r^{2}}e^{-\lambda}(r^{2}E)^\prime
\eea\ese
 for the line element (\ref{eq:1}). The quantity $\rho$ is the energy density, $p$ is the pressure, $E$ is the
 electric field intensity and $\sigma$ is the proper charge density. To integrate the system (\ref{eq:2}) it is necessary to choose two of the variables
 $\nu,~\lambda,~ \rho,~ p~ \textrm{or}~ E$. In our approach we specify $\lambda ~\textrm{and }E$.


 In the integration procedure, we make the
 choice\be\label{eq:3}
e^{2\lambda}=\frac{1-kr^{2}/R^{2}}{1-lr^{2}/R^{2}}\ee where $k$ and
$l$ are arbitrary constants. Note that the choice (\ref{eq:3})
ensures that the metric function $e^{2\lambda}$ is regular and
finite at the centre of the sphere. When $k=-7$ and $l=1$, in the
absence of charge, we regain the Tikekar interior metric \cite{Tik}
which models a superdense neutron star. Also observe that when $l=1$
we regain the metric function considered by Komathiraj and Maharaj
\cite{KoMa} which generalises the Maharaj and Leach \cite {MaLe} and
Tikekar \cite{Tik} models. Therefore particular choices of the
parameters $k$ and $l$ produce regular charged spheres which are
physically reasonable. Also the choice (\ref{eq:3}) ensures that
charged spheres generated, as exact solutions to the
Einstein-Maxwell system, contain well behaved uncharged models when
$E=0$.
 On eliminating $p$ from (\ref{eq:2b}) and
(\ref{eq:2c}), for the choice (\ref{eq:3}), the condition of
pressure isotropy with a nonzero electric field becomes \bea
\label{eq:4}
\left(1-kr^{2}/R^{2}\right)^{2}E^{2}&=&\left(1-kr^{2}/R^{2}\right)\left(1-lr^{2}/R^{2}\right)\left(\nu''+\nu'^{2}-\frac{\nu'}{r}\right) \nonumber\\
& & -(l-k)\frac{r}{R^{2}}\left(\nu'+\frac{1}{r}\right)+\frac{l-k}{R^{2}}\left(1-kr^{2}/R^{2}\right)
\eea
which is nonlinear.

To linearise the above equation it
is now convenient to introduce the transformation
\bse\label{eq:5}
\bea \psi(x)&=&e^{\nu}\\
x^{2}&=&1-lr^{2}/R^{2}
\eea\ese
 where $l\neq 0$. This transformation
 helps to simplify the integration procedure but
 changes the form of the potentials and matter variables. Then
(\ref{eq:4}) becomes \be \label{eq:6}
(l-k+kx^{2})\ddot{\psi}-kx\dot{\psi}+\left(\frac{(l-k+kx^{2})^{2}R^{2}E^{2}}{l^{2}(x^{2}-1)}+\frac{k(k-l)}{l}\right)
\psi=0\ee in terms of the new dependent and independent variables
$\psi$ and $x$ respectively. Equation (\ref{eq:6}) must be
integrated to find $\psi$, i.e. the metric function $\lambda$.
Note that the Einstein-Maxwell system (\ref{eq:2}) implies
\bse\label{eq:7}
\bea
\rho&=&\frac{l(l-k)}{R^{2}}\frac{(3l-k+kx^{2})}{(l-k+kx^{2})^{2}}-\frac{1}{2}E^{2}\\
p&=&\frac{l}{R^{2}(l-k+kx^{2})}\left(-2lx\frac{\dot{\psi}}{\psi}+k-l\right)+\frac{1}{2}E^{2}\\
\sigma^{2}&=&\frac{l^{2}}{R^{2}}\frac{[2xE-(1-x^{2})\dot{E}]^{2}}{(1-x^{2})(l-k+kx^{2})}
\eea\ese
 in terms of
 the variable $x$. Note that we have essentially reduced the
solution of the field equations to integrating (\ref{eq:6}). It is
necessary to specify the electric field intensity $E$  to complete
the integration. Only a few choices for $E$ are  physically
reasonable and generate closed form solutions. We can reduce
(\ref{eq:6}) to simpler form if we let \be \label{eq:8}
E^{2}=\frac{\alpha kl(x^{2}-1)}{R^{2}(l-k+kx^{2})^{2}}\ee where
$\alpha$ is constant. When $\alpha=0$ or $k=0$ there is no charge.
The form for $E^{2}$ in (\ref{eq:8}) vanishes at the centre of the
star, and remain continuous and bounded in the interior of the
star for a wide range of values of the parameters $\alpha,~k$ and
$l$. Upon substituting the choice (\ref{eq:8}) into (\ref{eq:6}),
we obtain \be \label{eq:9}
l(l-k+kx^{2})\ddot{\psi}-klx\dot{\psi}+k(k-l+\alpha)\psi=0\ee
which is the master equation for the system (\ref{eq:7}). We
expect that our investigation of equation (\ref{eq:9}) will
produce viable models of charged stars since the special cases
$\alpha=0$ and $\alpha \neq 0,~k\neq0,~l=1$ yields
models consistent with neutron stars.

\noindent
\bc{\bf 3. NEW SOLUTIONS}\ec

As the point $x=0$ is a regular point
of (\ref{eq:9}), there exists two linearly independent series
solutions with centre $x=0$. Thus we must have
 \be \label{eq:10}
\psi(x)=\sum_{i=0}^{\infty}a_{i}x^{i}\ee where $a_{i}$ are the
coefficients of the series. For an acceptable solution we need to
find the coefficients $a_{i}$ explicitly. On substituting
(\ref{eq:10}) in (\ref{eq:9}) we obtain after simplification
 \be \label{eq:11}
l(l-k)(i+1)(i+2)a_{i+2}+k[\alpha+k-l+li(i-2)]a_{i}=0,~ i \geq 0
\ee The equation  (\ref{eq:11}) is the basic difference equation
governing the structure of the solution. It is possible
 to express the general form for the even coefficients and odd coefficients in terms of the leading coefficient $a_{0}$ and $a_{1}$
  respectively by using the principle of mathematical induction. We generate a pattern \be \label{eq:12}
a_{2i}=\left(\frac{k}{l(k-l)}\right)^{i}\frac{1}{(2i)!}\prod_{q=1}^{i}[\alpha
+k-l+l(2q-2)(2q-4)]a_{0}\ee for the even coefficients
$a_{0},~a_{2},~a_{4}~.~.~.$.  Also we find the pattern
 \be \label{eq:13}
a_{2i+1}=\left(\frac{k}{l(k-l)}\right)^{i}\frac{1}{(2i+1)!}\prod_{q=1}^{i}[\alpha
+k-l+l(2q-1)(2q-3)]a_{1} \ee for the odd coefficients
$a_{1},~a_{3},~a_{5}~.~.~.$. Here the symbol $\prod$ denotes
multiplication.

From (\ref{eq:10}), (\ref{eq:12}) and (\ref{eq:13}), we can write
the general solution of  (\ref{eq:9}) as
\be \label{eq:14}
\psi(x)=a_{0}\psi_{1}(x)+a_{1} \psi_{2}(x)
\ee where we have set
\bse \label{eq:15}
\bea \label{eq:15a}
\psi_{1}(x)&=&
\left(1+
\sum_{i=1}^{\infty}\left(\frac{k}{l(k-l)}\right)^{i}\frac{1}{(2i)!} \times \right.  \nonumber \\
& & \left. \prod_{q=1}^{i}[\alpha
+k-l+l(2q-2)(2q-4)]x^{2i}\right)\\
\label{eq:15b}  \psi_{2}(x) &= &
 \left(x+
\sum_{i=1}^{\infty}\left(\frac{k}{l(k-l)}\right)^{i}\frac{1}{(2i+1)!} \times \right. \nonumber\\
 & & \left. \prod_{q=1}^{i}[\alpha
+k-l+l(2q-1)(2q-3)]x^{2i+1} \right).
\eea \ese
Thus we have
found the general solution to the differential equation
(\ref{eq:9}) for the particular choice of the electric field
(\ref{eq:8}). Series (\ref{eq:15a}) and (\ref{eq:15b}) converge if
there exists a radius of convergence which is not less than the
distance from the centre of the series to the nearest root of the
leading coefficient in (\ref{eq:9}). This is possible for a range
of values of $k$ and $l$.

The general solution (\ref{eq:14}) is given in the form of a
series which may be used to define new special functions. For
particular values of the parameters $\alpha,~k$ and $l$ it is
possible for the general solution to be written in terms of
elementary functions which is a more desirable form for the
physical description of a charged relativistic star. Solutions
that can be written in terms of polynomials and algebraic
functions can be found. This is a lengthy and tedious process and
we therefore  do not provide the details; the procedure is similar
to that presented in Komathiraj and Maharaj \cite{KoMa} which can
be referred to. The solutions found can also be verified with the
help of software packages such as Mathematica. Consequently we
present only the final solutions avoiding unnecessary details.

Two classes of solutions in terms of elementary functions can be
found. These can be written in terms of polynomials and algebraic
functions. The first category of solution for $\psi(x)$ is given
by
\bea \label{eq:16}
\psi(x)&=&A\sum_{j=0}^{n}(-\gamma)^{j}\frac{(n+j-2)!}{(n-j)!(2j)!}x^{2j}\nonumber \\
& & + B(l-k+kx^{2})^{3/2}\sum_{j=0}^{n-2}(-\gamma)^{j}\frac{(n+j)!}{(n-j-2)!(2j+1)!}x^{2j+1}
\eea
with the values
\beqas \gamma&=&4-\frac{4l}{4ln(n-1)+\alpha}\\
k+\alpha&=&l[2-(2n-1)^{2}]
\eeqas
The second category of solution
for $\psi(x)$ has the form
 \bea \label{eq:17}
\psi(x)&=&A\sum_{j=0}^{n}(-\mu)^{j}\frac{(n+j-1)!}{(n-j)!(2j+1)!}x^{2j+1}\nonumber \\
&& + B(l-k+kx^{2})^{3/2}\sum_{j=0}^{n-1}(-\mu)^{j}\frac{(n+j)!}{(n-j-1)!(2j)!}x^{2j}\eea
with the values
 \beqas \mu&=&4-\frac{4l}{4ln^{2}-l+\alpha}\\
k+\alpha &=& 2l(1-2n^{2})
\eeqas
 where $A$ and $B$ are arbitrary
constants and $x^{2}=1-lr^{2}/R^{2}$.

\noindent
\bc{\bf 5. SPECIAL CASES}\ec

From our general class of solutions  (\ref{eq:16}) and
(\ref{eq:17}), it is possible to generate particular cases found
previously. These can be explicitly regained directly from the
general series solution (\ref{eq:14}) or the elementary functions
(\ref{eq:16}) and (\ref{eq:17}). We demonstrate that this is
possible in the following classes.

We set $k+\alpha = -7l (n=2)$. Then $\gamma =
4(7l+\alpha)/(8l+\alpha) $ and it is easy to verify that equation
(\ref{eq:16}) becomes
 \beqas
\psi(x)&=&A'\left(1-4\left(\frac{7l+\alpha}{8l+\alpha}\right)x^{2}+\frac{8}{3}\left(\frac{7l+\alpha}{8l+\alpha}\right)^{2}x^{4}\right)\\
& & + B'x\left(1-\left(\frac{7l+\alpha}{8l+\alpha}\right)x^{2}\right)^{3/2}
\eeqas
where $A'=A/2$ and $B'=2B(8l+\alpha)^{3/2}$ are new
constants. Further setting $\alpha=0$ and $l=1$, we obtain \be
\psi(x)=A'\left(1-\frac{7}{2}x^{2}+\frac{49}{24}x^{4}\right)+B'x\left(1-\frac{7}{8}x^{2}\right)^{3/2}\n\ee
and $x^{2}=1-r^{2}/R^{2}$. Thus we have regained the Tikekar model
\cite{Tik} which is a viable model in the modelling of superdense
stars.

We set $k+\alpha = -2l (n=1)$. Then $\mu =
4(2l+\alpha)/(3l+\alpha)$ and  (\ref{eq:17}) becomes
 \beqas
\psi(x)&=&A''x\left(1-\frac{2}{3}\left(\frac{2l+\alpha}{3l+\alpha}\right)x^{2}\right)\\
&+&B''\left(1-\left(\frac{2l+\alpha}{3l+\alpha}\right)x^{2}\right)^{3/2}
\eeqas
 where are  $A''=A$ and $B''=B(3l+\alpha)^{3/2}$ are new
constants. Further setting $\alpha=0$ and $l/R^{2}=C/2~
(k/R^{2}=-C)$ and letting $X=Cr^{2}$ we obtain
\[ \tilde{\psi}=
\frac{A''}{9\sqrt{2}}(2-X)^{1/2}(5+2X)+\frac{B''}{3^{3/2}}(1+X)^{3/2}\]
 where we have set $\tilde{\psi}=\psi(X)$. Thus we have regained the Durgapal and Bannerji \cite{DuBa} model
  which is widely used in the modelling of neutron stars.

If we set $l=1$ and $\alpha=0$ then (\ref{eq:16}) and
(\ref{eq:17}) reduce to the corresponding expressions in the
solution of Maharaj and Leach \cite{MaLe} which implies a wide
family of models for uncharged relativistic spheres which have the
advantage of being expressed in elementary functions.

If we set $l=1$ then (\ref{eq:16}) and (\ref{eq:17}) contain the
solution of Komathiraj and Maharaj \cite{KoMa} for charged spheres
which are generalizations of earlier models with spheroidal
geometry.

\noindent
\bc{\bf 5. DISCUSSION}\ec

We have studied the Einstein-Maxwell system of equations for a
particular choice of the electric field intensity. The gravitational
potential was generalised to include the spheroidal geometry of the
hypersurfaces \textit{t}=constant of previous investigations. When
$l=1$ then we regain the Tikekar \cite{Tik} model and other exact
solutions found previously. We demonstrated that it was then
possible to reduce the condition of pressure isotropy to a second
order linear ordinary differential equation. This equation can be
solved in general using the method of Frobenius and the solution are
in terms of new special functions. Solutions in terms of elementary
functions can be extracted from the general solution for specific
parameter values. Particular models studied previously are contained
in our general class of solution. These solutions may be useful in
studying the physical behaviour of dense charged objects in
relativity which will be the objective in future work.

\begin{figure}[thb]
\vspace{1.5in} \includegraphics{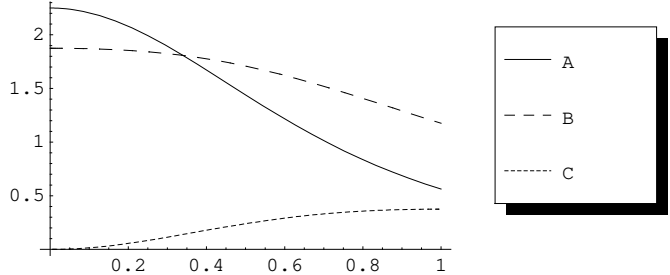}
 \vspace{2mm}
\caption{\label{plots1} Plots of the matter variables $\rho$, $p$
and $E^2$. \label{fig}}
\end{figure}

We briefly discuss the behaviour of the matter variables close to
the centre. We can graphically represent the the matter variables in
the stellar interior
 for particular choices of the parameter values. To this end
 we have produced Figure \ref{fig} with the
 help of the software package Mathematica.
 We have set $A=B=C=1$, $ k=- \frac14$, $l=-1$ and $\alpha = \frac32$
 over the interval $0\leq r\leq 1$,
 to generate the relevant plots in Figure \ref{fig}. Plots A and B denote
 the profiles of
energy density $\rho$ and the pressure $p$;
 plot $C$ denotes the electric field intensity $E^2$.
 We observe that these matter variables
 remain regular in the interior.
 We note that the energy density $\rho$ and the pressure $p$
 are positive and finite; they are monotonically decreasing functions
in the interior. The electric field intensity $E^2$   is  positive
and monotonically increasing in this interval. Thus the quantities
$\rho$, $p$ and $E^2$ are finite, continuous in the interval.

\noindent
\bc{\bf ACKNOWLEDGEMENTS}\ec

SDM acknowledges that this work is based upon research supported
by the South African Research Chair Initiative of the Department
of Science and Technology and the National Research Foundation.

\bc{\bf 5. REFERENCES}\ec
\vspace{-1.8cm}

\renewcommand{\refname}{}

\end{document}